\documentclass[aps,pre,10pt,amsmath,amssymb,twocolumn,showpacs,superscriptaddress,floatfix]{revtex4-1}
\usepackage{ifpdf}
 \newif\ifpdf
\ifx\pdfoutput\undefined
   \pdffalse
\else
   \pdfoutput=1
   \pdftrue
\fi
\ifpdf
   \usepackage{graphicx}
\else
   \usepackage{graphicx}
\fi
\usepackage{hyperref}

\DeclareMathOperator{\SP}{\mathrm{\scriptscriptstyle SP}}
\DeclareMathOperator{\QD}{\mathrm{\scriptscriptstyle QD}}
\DeclareMathOperator{\NP}{\mathrm{\scriptscriptstyle NP}}
\DeclareMathOperator{\Ra}{\mathrm{\scriptscriptstyle R}}
 \DeclareMathOperator{\tr}{tr}

\begin{document}

\title{Nanolaser in the selfgenerated nonequilibrium environment: quantum fluctuations and entanglement}

\author{E.~S.~Andrianov}
\affiliation{Institute for Theoretical and Applied Electromagnetics, 13 Izhorskaya, Moscow 125412, Russia}
\affiliation{Department of Theoretical Physics, Moscow Institute of Physics and Technology, 141700 Moscow, Russia}

\author{N.~M.~Chtchelkatchev}
\affiliation{Department of Theoretical Physics, Moscow Institute of Physics and Technology, 141700 Moscow, Russia}
\affiliation{Institute for High Pressure Physics, Russian Academy of Science, Troitsk 142190, Russia}
\affiliation{Department of Physics and Astronomy, California State University Northridge, Northridge, CA 91330, USA}

\author{A.~A.~Pukhov}
\affiliation{Institute for Theoretical and Applied Electromagnetics, 13 Izhorskaya, Moscow 125412, Russia}
\affiliation{Department of Theoretical Physics, Moscow Institute of Physics and Technology, 141700 Moscow, Russia}

\begin{abstract}
We investigate the dynamics of the spaser-based nanolaser in the strong incoherent pumping regime in the quantum limit when the photon number is the order of unity. We consider the situation where the newly irradiated photon finds itself in the cloud of earlier irradiated photons that are not thermalized. As the result the entanglement of nanoparticle with quantum dot degrees of freedom in the nanolaser and the lasing intensity increases several times. In fact the nonthermal bath effectively makes the nanolaser ``more quantum'' and master equation for the nanolaser density matrix nonlinear and selfconsistent.
\end{abstract}

\maketitle


Coherent manipulation and entanglement of photon and condensed matter quantum degrees of freedom is the topic of intensive discussions with possible applications for quantum calculations~\cite{niemczyk2010circuit,astafiev2007single}. Recent discoveries in the area of nanoplasmonics have raised high hopes for the future development of ultrafast and super small quantum optoelectronic devices~\cite{gaponenko2010introduction,maier2007plasmonics,kawata2007tip,wang2010nanoscale}. The key element of an active plasmonic devise is the spaser based nanolaser~\cite{bergman2003surface,noginov2009demonstration,stockman2010spaser,andrianov2012stationary}.

The nanolaser is the nanosystem coherent, entangled and dissipative at the same time~\cite{strauf2011single}. The spaser consists of a quantum dot (QD) placed near a metallic nanoparticle (NP). The physical principle of the spaser’s operation is similar to that of laser. The role of photons is played by surface plasmons (SPs), which are localized at the NP. Confining SPs to the NP resembles a resonator. The spaser generates and amplifies the near field of the NP. SP amplification occurs because of nonradiative energy transfer from the QD to the NP. This process originates from the dipole-dipole (or any other near-field) interaction between the QD and the plasmonic NP.  The generation of a large number of SPs leads to the induced emission of the QD into the plasmonic mode and to the development of generation of plasmons. Thus, the excitation of the plasmonic mode is provided by pumping through the excited QD. This process is inhibited by losses in the NP, which together with pumping results in undamped stationary oscillations of the spaser dipole moment.

\begin{figure}[b]
  \centering
  \includegraphics[width=0.95\columnwidth]{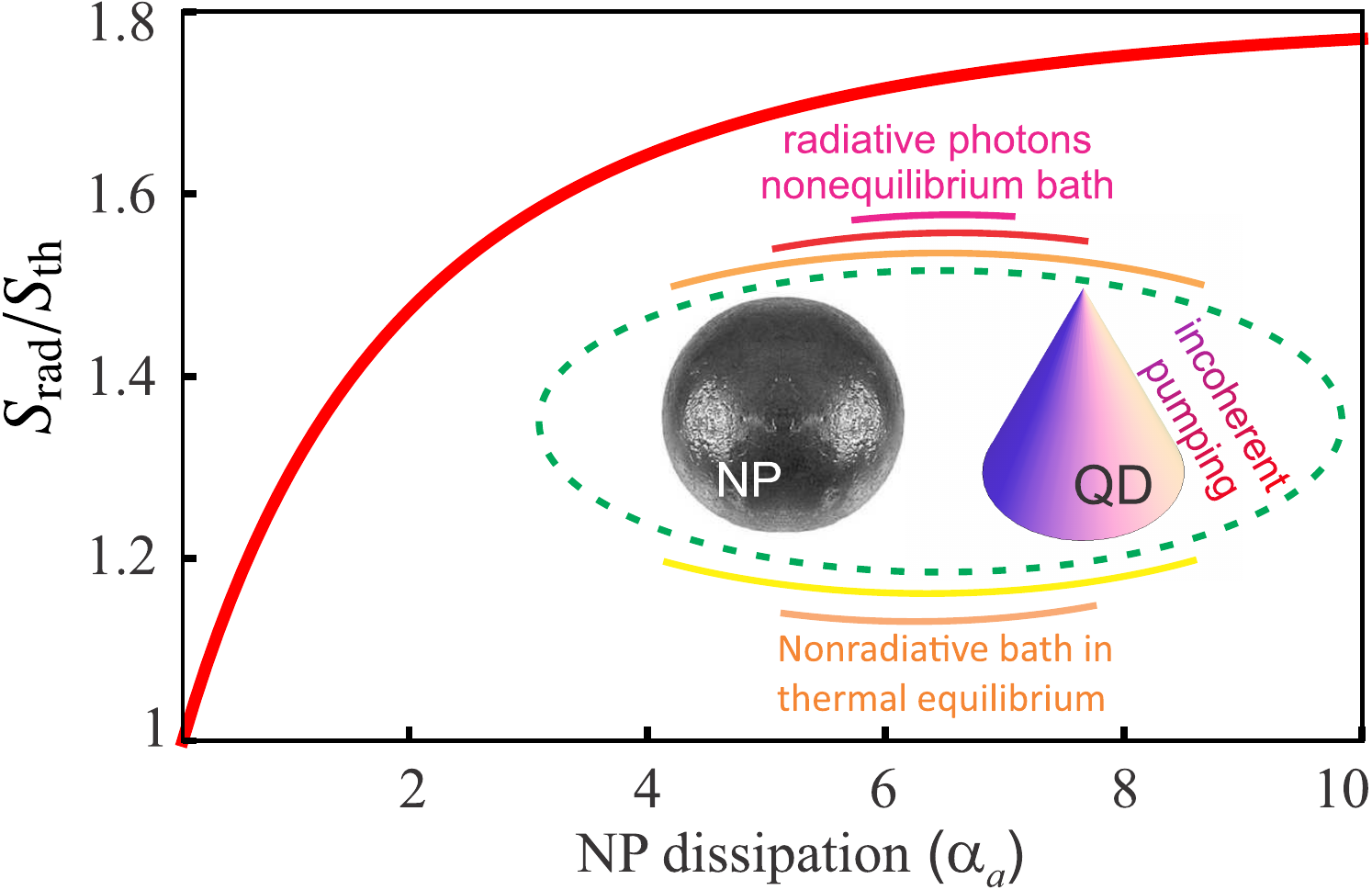}\\
  \caption{(Color online) The nonthermal bath effectively makes the nanolaser ``more quantum'': the graph shows the ratio of the entanglement entropies $S_{\rm rad}/S_{\rm th}>1$ describing the entanglement of plasmons and quantum dot degrees of freedom in the nonequilibrium and thermal environment correspondingly as the functions of the dissipation due to NP radiation. Here $\alpha_a$  is the ratio of the thermal and nothermal relaxation times for nanoparticle; $\alpha_a=0$ corresponds to linear Lindblad theory and thermal bath. Making the graph we kept the ratio of the thermal and nothermal relaxation times for the the quantum dot $\alpha_\sigma=0$. Inset: energy balance diagram of the nanolaser. The distribution of photons radiated by the nanolaser is nonthermal and they effectively form the nonequilibrium photon bath for the nanolazer.}\label{fig1}
\end{figure}
There are two channels for energy dissipation in the nanolaser. The first one is standard: it is phonon energy relaxation, Joule losses and other nonradiative processes~\cite{bergman2003surface,noginov2009demonstration,stockman2010spaser,andrianov2012stationary,waks2010cavity}. The second channel is related to the irradiation and absorbtion of photons, see Fig.~\ref{fig1}. The dissipation of energy in this channel is regulated by the density matrix of the photon bath. There are thermal photons that come from the thermal photon bath (this photons come from afar). Other photons are generated by the nanolaser itself and they have the nonthermal distribution determined by the quantum state of the nanolaser. Moreover the distribution of emitted photons is determined by quantum dynamics of plasmon degrees of freedom. These irradiated photons thermalize far from the nanolazer. The nanoloser is typically placed into the (nonuniform) dielectric matrix that in turn can be placed into the resonator-box. We concentrate on the situation when the irradiated photons do not immediately leave the nanolaser but may return back or stay nearby after the reflections on the inhomogeneities of the dielectric matrix, the interfaces of the dielectric layers \textit{etc}. These photons effectively form the nonequilibrium photon bath for the nanolaser.

The QD-transition frequency and the frequency of plasmonic resonance are typically larger than the temperature of the thermal bath so the average thermal photon number is much smaller than unity. However the average number of the nonthermal photons is typically of the order of unity or larger depending on pumping. That suggests the strong two-way influence of the nonequilbruim photon bath on the dynamics of the nanolaser, see Fig.~\ref{fig1}. When the new photon is irradiated by the nanolaser it finds itself in the cloud of earlier irradiated photons (nonequilibrium photon bath). The occupation numbers of the photons in the bath influence on the rate of the irradiation and so on the nanolaser quantum state.

Intuitively when we drive the quantum system by some bath it results in the suppression of the quantum correlations (entanglement) between the system degrees of freedom. The bath typically results in the random forces with the dispersion related to the effective temperature of the bath~\cite{Kamenev2005,carmichael1993open}. That thermal forces destroy entanglement~\cite{nielsen2010quantum}. In our case this simple picture does not work: our bath is essentially nonthermal (our quantum system in fact produces the bath for itself) that effectively results in the nonlinear coupling of the bath with the quantum system.

We show below that the nonequilibrium nature of the thermostat changes the quantum physics of the nanolaser. It induces in particular strong (nonlinear) dependence of the plasmon number and the degree of the QD-NP entanglement on the dissipation rates. We demonstrate that the nonequilbrium nature of the photon bath makes the dissipation in the master equation for the nanolaser density matrix selfconsistently dependent on the occupation number distribution of emitted photons. So the master equations becomes nonlinear. As the result quantum fluctuations strongly increase. In fact the nonthermal bath effectively makes the nanolaser ``more quantum'' than that with the thermal bath only.

Standard approach for description of an open strong dissipative quantum system like nanolaser is the master equation approach for the reduced density matrix~\cite{scully1997quantum,carmichael1993open,walls2008quantum}. In usually used Linblad master equations the dissipation is related with the thermal bath. In our case the irradiation becomes effectively the nonequilibrium bath with the state determined by the coherent quantum dynamics of the nanolaser. The possible way to resolve the coexistence of strong dissipation and quantum physics in nanolaser far from equilibrium is the nonlinear master equation for the nanolaser density matrix.

The master equation for reduced density matrix has the form~\cite{scully1997quantum,carmichael1993open,walls2008quantum}
\begin{gather}\label{eq1}
 \frac{d}{dt} \hat \rho=-i[\hat H,\hat \rho]+\mathcal{\hat L} \hat \rho,
\end{gather}
where $\hat\rho$ is the reduced density matrix, $\mathcal{\hat L} $ is Lindbladian superoperator, and $\hat H_0$ is system Hamiltonian. Lindbladian describes the dissipation and typically it does not depend on the system state. This approach works well for thermal reservoirs. But in our case the irradiation plays the role of nonthermal (nonequilibrium) reservoir. We can take the nonequilibrium bath into account making Eq.~\eqref{eq1} effectively nonlinear. Typically the information about the bath is encoded in the $c$-number parameters of the Lindbladian. In our case the state of the irradiative bath depends on the reduced density matrix $\hat \rho$. So the $c$-number parameters of the Lindbladian responsible for the irradiation are related to $\hat \rho$. They can be considered as certain averages of the appropriate system operators with $\hat\rho$. Thus the Lindbladian selfconsistently depends on $\hat \rho$.

Using this approach we have found the nonlinear dynamics of plasmon and as a result the stationary value of the average plasmon number is by several times larger than that found from the linear Lindbladian equation. We calculated entanglement entropy and got that the degree of entanglement between plasmon and quantum dot strongly increases near generation threshold. We predict that registering the nonequilibrium irradiation of photons we can make judgements about the quantum  state  of the nanolaser.


The Hamiltonian of for the surface plasmons $\hat H_{\SP}=\hbar\omega_{\SP}\hat a^\dag a$, where $\omega_{\SP}$ is the plasmon frequency and $\hat a^\dag$ is the plasmon creation operator in the dipole mode. The Hamiltonian of the QD $ \hat H_{\QD}=\hbar\omega_{\QD}\hat\sigma^\dag\hat\sigma$, where $\omega_{\QD}$ is the transition frequency between the levels being in resonance with the SP, and $\hat\sigma$ is the operator of the transition between the excited $|e\rangle$  and ground $|g\rangle$ states of the QD. The interaction between SP and QD we take in the  Janes-Cummings form: $\hat H_i=\hbar\omega_{\Ra} \hat a^\dag\hat\sigma+h.c.$, where the coupling constant $\omega_{\Ra}$ is the Rabi frequency. So Hamiltonian of the spaser $\hat H=\hat H_{\SP}+\hat H_{\QD}+\hat H_i$~\cite{strauf2011single,stockman2011nanoplasmonics,vinogradov2012quantum}.

The Lindblad superoperator acts on the density matrix as follows: $\mathcal{\hat L}\hat\rho=\sum_j(\hat V_j\hat\rho \hat V_j^\dag-\frac12\{\hat\rho,\hat V_j^\dag\hat V_j\})$, where $\hat V_j$ are transition operators. We will distinguish two types of reservoirs. The first one is the bath in thermal equilibrium. For the atomic system there two types of the transition operators corresponding to damping to this reservoirs are:
\begin{gather}\label{Vsigma}
\hat V_{\sigma_{\rm dump1}}^{\rm th}=\hat \sigma\sqrt{\frac{1+N_{\sigma}^{\rm th}}{\tau_{\sigma}^{\rm th}}},\qquad \hat V_{\sigma_{\rm dump2}}^{\rm(th)}=\hat \sigma^\dag\sqrt{\frac{N_{\sigma}^{\rm th}}{\tau_{\sigma}^{\rm th}}},
\end{gather}
where $\tau_{\sigma}^{\rm th}$ is the damping rate. Here $N_{\sigma}^{\rm th}$ is the average number of quanta in the thermal bath at the QD transition frequency.
Similarly we write the transition operators acting on the plasmon degrees of freedom
\begin{gather}
  \hat V_{a_{\rm dump1}}^{\rm th}=\hat a\sqrt{\frac{2[1+N_{a}^{\rm th}]}{\tau_a^{\rm th}}},\qquad  \hat V_{a_{\rm dump2}}^{\rm th}=\hat a^\dag\sqrt{\frac{2N_{a}^{\rm th}}{\tau_a^{\rm th}}},
\end{gather}
where $\tau_{a}^{\rm th}$ is the dumping rate of nanoparticle dipole moment and $N_{a}^{\rm th}$ is the average number of quanta in the thermal bath at the frequency of plasmonic resonance. At optical frequency and room temperature $N_{\sigma}^{\rm th}=N_{a}^{\rm th}=0$.
To consider incoherent pumping, following Ref.~\cite{carmichael1993open} we describe pumping using the Lindblad superoperator:
\begin{gather}
\hat V_{\sigma_{\rm pump}}=\hat \sigma^\dag\sqrt{\frac{1}{\tau_p}},
\end{gather}
where $\tau_{p}$ is the pumping rate.

\begin{figure*}[t]
  \centering
  \includegraphics[width=0.99\columnwidth]{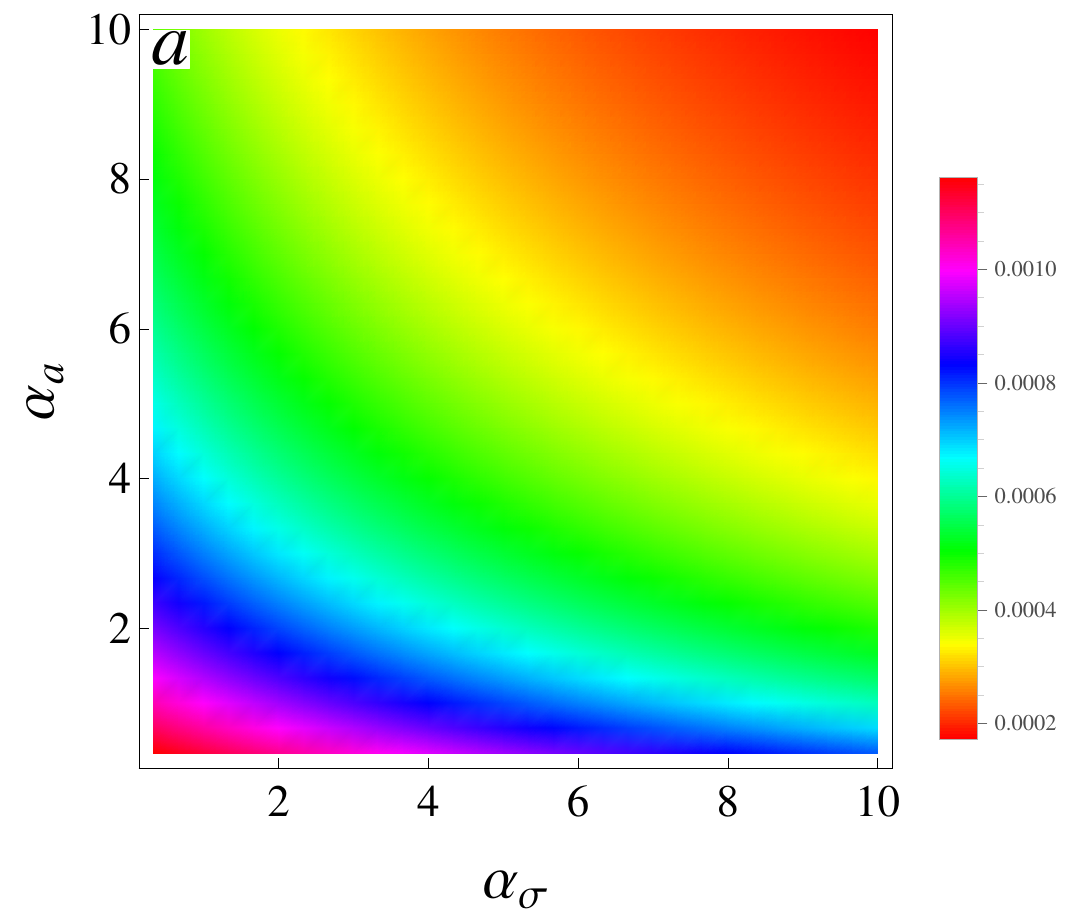} \includegraphics[width=0.99\columnwidth]{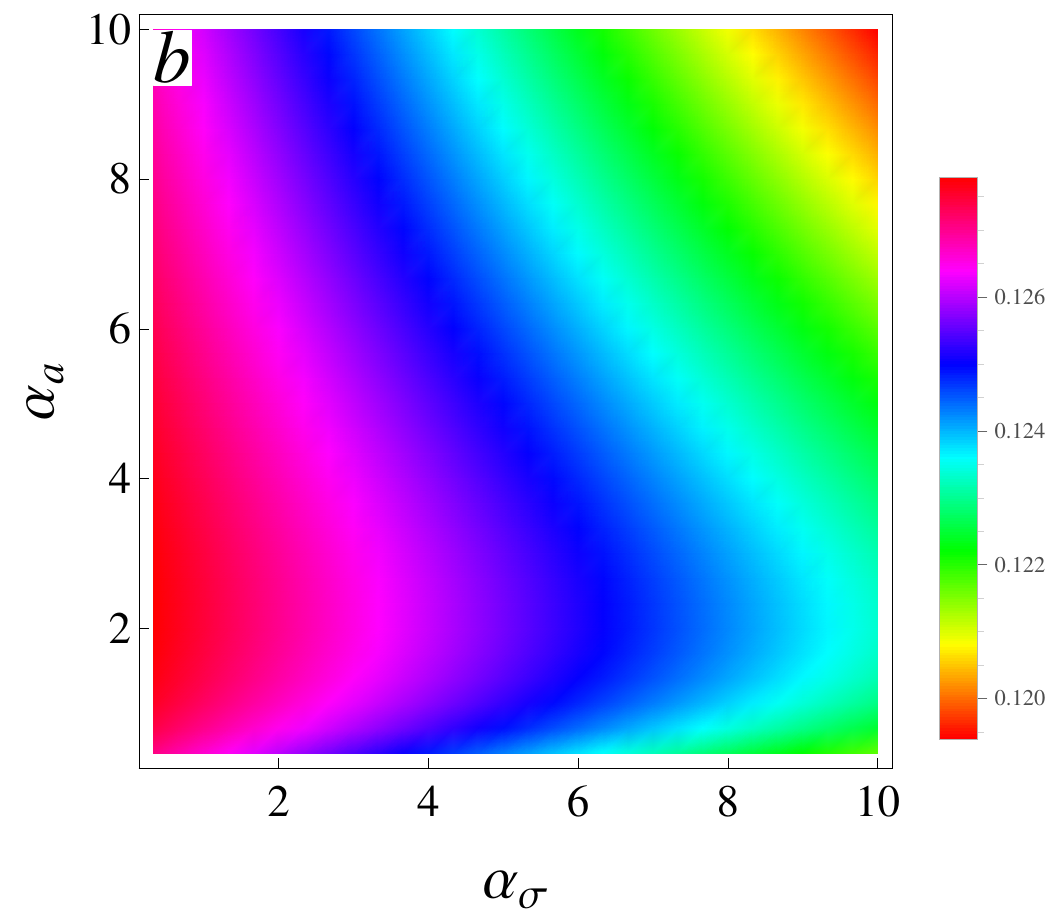}
  \\
  \includegraphics[width=0.99\columnwidth]{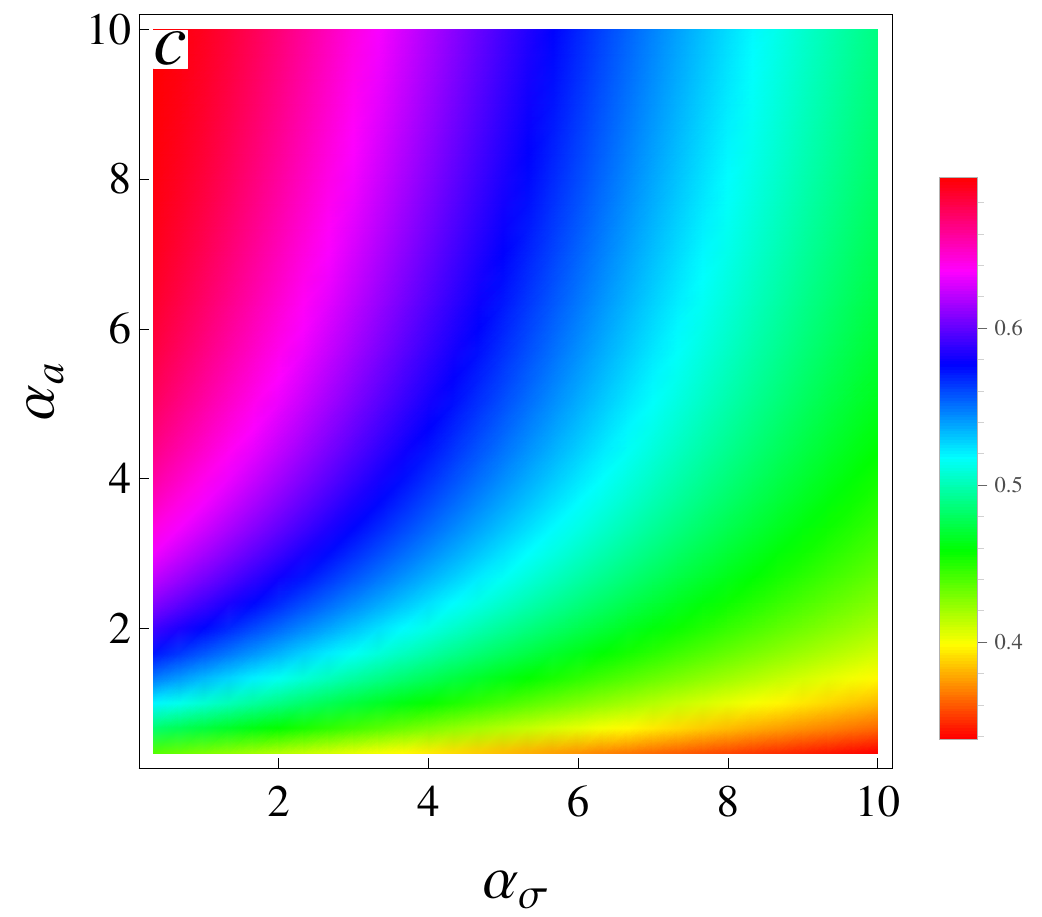} \includegraphics[width=0.99\columnwidth]{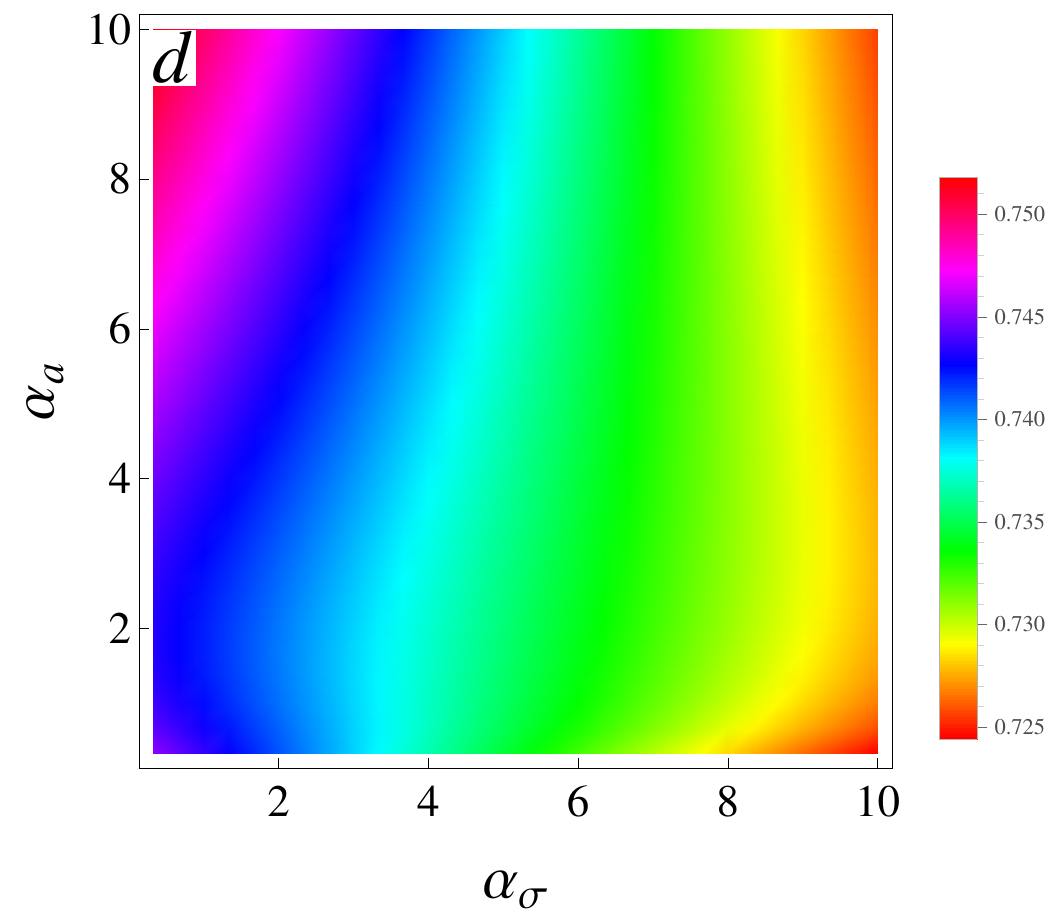}
  \caption{(Color online)  The dependence of plasmon number (a,b) and entropy (c,d) on the dissipation in the two channels: due to radiation for NP ($\alpha_a$) and QD ($\alpha_\sigma$) respectively. Left and right figures correspond to $\tau^{\rm th}_a=10^{-14}s$, $\tau^{\rm th}_\sigma=10^{-11}s$, $\tau_p=10^{-12}s$ and $\tau^{\rm th}_a=10^{-12}s$, $\tau^{\rm th}_\sigma=10^{-12}s$, $\tau_p=10^{-12}s$ (we took typical experimental parameters). We remind, $\alpha_a$ ($\alpha_\sigma$) is the ratio of the thermal and nothermal relaxation times for nanoparticle and quantum dot respectively; $\alpha_a=0$ and $\alpha_\sigma=0$ corresponds to linear Lindblad theory and thermal bath. It should be noted that the color gradients in the density plots of the plasmon number and entropy here are induced by the nonequilibrium dissipation. }\label{figdensity1}
\end{figure*}
The second type of reservoir is connected to the irradiation of NP and QD dipole moments. It consists of the cloud irradiated nonequilibrium photons near the spaser. Formally, the Lindblad operators for this type of reservoirs are analogous to ones for thermal bath:
\begin{gather}
\hat V_{\sigma_{\rm dump1}}^{\rm rad}=\hat \sigma\sqrt{\frac{1+N_{\sigma}^{\rm rad}}{\tau_{\sigma}^{\rm rad}}},\qquad \hat V_{\sigma_{\rm dump2}}^{\rm rad}=\hat \sigma^\dag\sqrt{\frac{N_{\sigma}^{\rm rad}}{\tau_{\sigma}^{\rm rad}}},
\end{gather}
for QD and
\begin{gather}\label{Vdump}
  \hat V_{a_{\rm dump1}}^{\rm rad}=\hat a\sqrt{\frac{[1+N_{a}^{\rm rad}]}{\tau_a^{\rm rad}}},\qquad  \hat V_{a_{\rm dump2}}^{\rm rad}=\hat a^\dag\sqrt{\frac{N_{a}^{\rm rad}}{\tau_a^{\rm rad}}},
\end{gather}
for metal NP.

Lindbladian superoperators given above are more or less standard. But now we take into account that there are two types of reservoirs for dissipation of energy in nanolaser (spaser) and the state of our nonequilibrium photon reservoir selfconsistently depends on the nanolaser state. The electric filed operator is proportional to the dipole moment that in turn is linear operator over the creation (annihilation ) plasmon operators. So the average number of nonequilibrium photons should be proportional to the average number of surface plasmons on metal NP. The coefficient of proportionality is the model constant: it is fixed by the matrix elements of the dipole moment. For simplicity we take unity for this constant: qualitatively, all our results are stable with the respect to the choice of this constant. More general situation we would consider in~\cite{future}. Then the average number of nonequilibrium photons should be the same as the average number of surface plasmons on metal NP:
\begin{gather}\label{Nsigma}
  N_{\sigma}^{\rm rad}=N_{a}^{\rm rad}=\langle \hat a^\dag\hat a\rangle\equiv\tr\left\{\hat\rho[N_{a,\sigma}^{\rm rad}]\hat a^\dag\hat a\right\}.
\end{gather}
So now the Lindbladian depends on the density matrix,
\begin{gather}
 \mathcal{\hat L}=\mathcal{\hat L}[\hat \rho],
\end{gather}
and the master equation~\eqref{eq1} becomes nonlinear. This is one of our key statements.

Here we assume that the dipole moment of NP is much larger that one of the QD. So the nanoparticle mostly irradiates. In more general situation one can get  in Eq.~\eqref{Nsigma} the superposition of $\langle\hat\sigma_z\rangle$ and $\langle \hat a^\dag\hat a\rangle$. Similarly one can consider the situation when some fraction of the irradiation returns back to the nanolaser~\cite{future}.

We can understand the origin of nonlinearity in our selfconsistent Lindblad equations writing the energy of the electromagnetic field where we explicitly distinguish the electric field generated by ``cold'' photons from the thermal bath and the electric field generated by the irradiated ``hot'' photons:
\begin{gather}
  \int d^3r\frac{(\mathbf E_{\rm hot}+\mathbf E_{\rm cold})^2}{8\pi},
\end{gather}
where we have used the superposition principle. Then it follows that the interaction Hamiltonian looks like
\begin{gather}
  \int d^3r\frac{\mathbf{\hat E}_{\rm hot}\cdot \mathbf{\hat E}_{\rm cold}}{4\pi}.
\end{gather}

Each electric field operator can be expanded in the standard way over the creation and annihilation operators corresponding to the thermal bath and cloud of hot irradiated photons.
Then the interaction
\begin{gather}
  \hat H_{\rm int}\sim\sum_k (a_{k,\rm hot}a_{k,\rm cold}^\dag)+h.c.,
\end{gather}
where $\langle a^\dag_{k,\rm cold}a_{k',\rm hot}\rangle=0$ and $\langle a^\dag_{k,\rm cold(hot)}a_{k',\rm cold (hot)}\rangle=\delta_{k,k'} n_k^{\rm(cold(hot))}$, where $n_k^{\rm cold(hot)}$ is the distribution function of cold (hot) photons. We identify above $N^{\rm th(rad)}$ with the average occupation numbers of cold (hot) photons.

In fact what we have done with the Lindblad equations is similar to the Landauer-Buttiker approach~\cite{landauer1987electrical,PhysRevLett.57.1761,lesovik1989pis,de1994doubled,nagaev1992shot,martin1992wave} in quantum transport theory of mesoscopic nanocircuits. There observables describing transport phenomena (like the current operator) are expanded over the second-quantization creation-annihilation electron operators that correspond to different electron reservoirs that are not generally in equilibrium with each other (even not necessary equilibrium themselves), see, e.g., Refs.~\cite{blanter2000shot,nazarov1999novel} for a review. The core of the Landauer-Buttiker approach is having electron density matrix in the interaction picture being the direct product of the reservoir density matrices.  Electrons coming into the nanowire from the ``cold'' reservoir have one sort of creation-annihilation operators while the electrons coming to this nanowire from the ``hot'' reservoir are described by another set of creation-annihilation operators. In our case we have thermal bath (cold reservoir) and irradiation (hot reservoir).

The nonequilibrium nature of the thermostat strongly influences on the quantum physics of the nanolaser.  First of all it induces the (nonlinear) dependence of the plasmon number and the degree of the QD-NP entanglement (reduced entropy) on the dissipation due to NP ($\alpha_a$) and QD ($\alpha_\sigma$) radiation. If we have equilibrium thermostat then the plasmon number and the reduced entropy are nearly independent on the dissipation channels.

The numerical simulation has been performed by truncated Hilbert space of the nanolaser on 100 plasmons. This assumption is confirmed by the fact that the average number of plasmon is the order of unity and the accuracy of this truncation is much more than enough.

We consider below the illustrative examples that show the importance of the nonequilibrium two-channel dissipation. The parameters we choose are typical for experiment, see, e.g., Ref.~\cite{noginov2009demonstration}. We focus on the system with the positive pumping and different ratio of thermal relaxation times of NP ($\tau^{\rm th}_a$) and QD ($\tau^{\rm th}_\sigma$) and investigate the influence of the radiation losses on plasmon number and entropy. For convenience we introduce the following notations: $\alpha_a=\tau^{\rm th}_a/\tau^{\rm rad}_a$ and $\alpha_\sigma=\tau^{\rm th}_\sigma/\tau^{\rm rad}_\sigma$.

The first case corresponds to high Joule losses in metallic NP ($\tau^{\rm th}_a = 10^{-14}s$) and small thermal losses in QD ($\tau^{\rm th}_\sigma = 10^{-11}s$), the pumping value is less than the threshold value. In such situation the plasmon number decreases due to radiation because the loss intensity due to radiation is proportional to the plasmon number. But the entropy  increases due to radiation because the nonequilibrium thermal bath is common for NP and QD and they interact with each other through this reservoir (fig.1). It should be noted that maximum of the entropy corresponds to strong dissipation in the NP-channel while the strong QD-radiation is not accompanied by the increase of the entropy. This is because the nonequilibrium bath is largely formed by the NP-radiation rather than the QD-one. The entanglement entropy is determined by the formula $S=-\tr(\hat \rho_{\NP} \rm \ln \hat \rho_{\NP})$ where $\hat \rho_{\NP}=\tr_{\QD}(\hat \rho)$ according to~\cite{preskill1998lecture}.

The second case corresponds to comparable value of NP and QD thermal losses ($\tau^{\rm th}_a=\tau^{\rm th}_\sigma = 10^{-12}s$), the pumping value is slightly more than the threshold value. In this situation there is the value of NP radiation losses in which the plasmon number achieves maximum value. At the same time the entropy in such radiation losses has a minimum. Further increase of NP radiation losses causes decrease of plasmon number and increase of entropy, Fig.~\ref{figdensity1}. In all cases QD radiation causes decreases both plasmon number and entropy.

To, conclude, we investigate the dynamics of the spaser based nanolaser in the nonequilibrium selfgenerated environment where the newly irradiated photon finds itself in the cloud of earlier irradiated photons. The nonthermal bath effectively makes the nanolaser ``more quantum'' as follows from the entanglement entropy. We obtain self-consistent nonlinear quantum master equation for the nanolaser density matrix which includes the influence of both thermal and non-thermal reservoirs. We find that nonequilubrium and nonlinear effects result in increase of number of plasmons to the order of magnitude (the same applies for photon intensity).

We thank A. Vinogradov and Yu. Lozovik for helpful discussions. The work was funded by RFBR No.~13-02-91177, 13-02-00579, 13-02-00407, NSF Grant DMR 1158666, Dynasty foundation, the Grant of President of Russian Federation for support of Leading Scientific Schools No.~6170.2012.2, RAS presidium and Russian Federal Government programs.

\appendix
\section*{Supplementary material}

\subsection{The set of main equation}
We wrote in the manuscript the Lindblad equations in the superoperator form. Here we write down the set of Lindblad equations explicitly:
\begin{widetext}\begin{multline}
  \frac{d}{dt} \hat \rho=-i[\hat H_{\rm JCM},\hat \rho]_-+\frac2{\tau^{\rm rad}_a}[1+\langle \hat a^\dag\hat a\rangle]\left(\hat a\hat \rho\hat a^\dag-\frac12\{\hat \rho,\hat a^\dag \hat a\}_+\right)+
 \frac2{\tau^{\rm rad}_a}\langle \hat a^\dag\hat a\rangle\left(\hat a^\dag\hat \rho\hat a-\frac12\{\hat \rho, \hat a\hat a^\dag\}_+\right)+
 \\
 \frac2{\tau^{\rm th}_a}[1+\langle n_{\rm th}\rangle]\left(\hat a\hat \rho\hat a^\dag-\frac12\{\hat \rho,\hat a^\dag \hat a\}_+\right)+
 \frac2{\tau^{\rm th}_a}\langle n_{\rm th}\rangle\left(\hat a^\dag\hat \rho\hat a-\frac12\{\hat \rho, \hat a\hat a^\dag\}_+\right)+
 \\
 \frac1{\tau^{\rm rad}_D}[1+\langle \hat a^\dag\hat a\rangle]\left(\hat \sigma_-\hat \rho\hat \sigma_+-\frac12\{\hat \rho,\hat \sigma_+\hat \sigma_-\}_+\right)+
 \frac1{\tau^{\rm rad}_D}\langle \hat a^\dag\hat a\rangle\left(\hat \sigma_+\hat \rho\hat \sigma_--\frac12\{\hat \rho,\hat \sigma_-\hat \sigma_+\}_+\right)+
 \\
 \frac1{\tau^{\rm th}_D}[1+\langle n_{\rm th}\rangle]\left(\hat \sigma_-\hat \rho\hat \sigma_+-\frac12\{\hat \rho,\hat \sigma_+\hat \sigma_-\}_+\right)+
 \frac1{\tau^{\rm th}_D}\langle n_{\rm th}\rangle\left(\hat \sigma_+\hat \rho\hat \sigma_--\frac12\{\hat \rho,\hat \sigma_-\hat \sigma_+\}_+\right)+
 \\
 \frac2{\tau_s}\left(\hat \sigma_z\hat \rho\hat \sigma_z-\hat \rho\right)+
 \frac1{\tau_p}\left(\hat \sigma_-\hat \rho\hat \sigma_+-\frac12\{\hat \rho,\hat \sigma_+\hat \sigma_-\}_+\right).
\end{multline}\end{widetext}
This nonlinear Lindblad equation is generalization of the linear one. The first term corresponds to the interaction between NP and QD. The second and third terms and the sixth and the seventh ones describe the dissipation due to the nonequilibrium photon bath of NP and QD respectively. Here we assume that $\langle n_{\rm rad}\rangle=\langle \hat a^\dag\hat a\rangle$. More accurate justification of this assumption is given in the main text. The fourth and the fifth terms and the eighth and the ninth ones correspond to the standard thermal reservoir dissipation. Finally, the penultimate term describes the phase destroying processes and the last one corresponds to the incoherent pumping~\cite{carmichael1993open}.

\subsection{Discussion}

We write in the paper that registering the nonequilibrium irradiation of photons we can make judgements about the quantum state of the nanolaser. Below we make this claim more quantitative. One can detect these nonequilibrium photons experimentally by the photodetectors and find the spectrum of photons, their average photon number and photon number cumulants.  This information would allow making judgments about the quantum state of the nanolaser, for example, from the average number of photons that is proportional to the average number of plasmons, because of the irradiated photon number $\left\langle \hat{a}^{+} \hat{a}\right\rangle $ is proportional to the nanoparticle dipole moment intensity. From the photon distribution function one can extract information about the plasmon distribution. Interference experiments can detect the non-diagonal elements of the photon density matrix. Then one can try to detect the photon entropy.

Below we comment why the average number of nonequilibrium photons should be the same as the average number of surface plasmons on metal NP. The electric filed operator is proportional to the dipole moment that in turn is linear operator over the creation (annihilation ) plasmon operators. So more accurate to say that the average number of nonequilibrium photons should be proportional the average number of surface plasmons on metal NP. The coefficient of proportionality is the model constant: it is fixed by the matrix elements of the dipole moment.

We give analogy in the paper between Landauer-Buttiker formalism for quantum electronic transport and our approach. Initially the Landauer-Buttiker formalism was formulated in such a way that both contacts were in equilibrium (though not in equilibrium with each other), while in our work we have an equilibrated bath of phonons and nonequilibrium bath of photons. Last time the Landauer-Buttiker formalism was significantly developed, see Refs.~[27-28]. It is not necessary now in the Landauer-Buttiker formalism having both terminals equilibrium.  The most important is having electron density matrix in the interaction picture being the direct product of the terminal density matrices.  For example, consider the following structure: Left terminal -- quantum wire -- quantum dot --- quantum wire -- Right terminal.  When the quantum dot is large enough (coulomb blockade is negligible), the quantum dot can be considered as the nonequilibrium bath with the electron distribution function being the linear combination of electron distribution functions in the leads, see the review of Blanter and Buttiker.  The idea to consider the quantum dots as the nonequilibrium bath in the Landauer-Buttiker formalism resulted in the quantum ``circuit theory'' theory of transport in quantum circuits, see, e.g.,~[28] and the review~[27].

We discuss in the text the ``pumping value'' and the ``threshold value''  and say that ``the pumping value is slightly more than the threshold value''.  From Lindblad equation we can obtain the equation for the value of incoherent pumping $D_{0} =\frac{\tau _{\sigma } /2-\tau _{p} }{\tau _{\sigma } /2+\tau _{p} } $ and from the semi-classical theory the pumping threshold $D_{\rm th} =1/\omega _{R}^{2} \tau _{a} \tau _{\sigma } $ corresponding to generation of coherent plasmon. So if the pumping intensity $\tau _{p} $ such that $D_{0} <D_{\rm th} $ we write ``that the pumping value is less than the threshold value'' and if the pumping intensity $\tau _{p} $ such that $D_{0} >D_{\rm th} $ we write ``the pumping value is more than the threshold value''.

It is known that the Master equation in the Lindblad form is derived within the following approximations. The first approximation is a consequence of the weak-coupling assumption, which allows one to expand the exact equation of motion for the density matrix to the second order with respect to the interaction between the system and the reservoir. Together with the decomposition of the density matrix into the system and reservoir parts this leads to the Born approximation to the master equation. Below we clerify how this assumption is adequate for the system with the strong coupling regime between the NP and QD dipole moments and the second type of reservoir, i.e. a cloud of the irradiated nonequilibrium photons near the spacer. Weak-coupling regime assumes that the interaction between system and reservoir is much less than the frequency of system oscillation. Since the nanolaser operates at optical frequency, $\omega_{\SP} \sim 10^{15} s^{-1} $, the Born approximation requires that the interaction constant is much less than $\omega_{\SP} $. It is true for nanolaser interacting with the thermal bath~\cite{waks}. In our case, the average number of irradiated photons does not far exceed unity, so our effective interaction constant (bare interaction constant times the average number of photons) is still much less than $\omega_{\SP} $ as we have for the case of the equilibrium reservoir. And the Born approximation remains true.

The second approximation is the Markov approximation in which the quantum master equation is made local in time by replacing the system density matrix at the retarded time with that at the present time. The relevant physical condition for this approximation is that the reservoir correlation time is small compared to the relaxation time of the system. Below we explain how this approximation is in accord with our statement that ``When the new photon is irradiated by the nanolaser it finds itself in the cloud of earlier irradiated photons. The occupation numbers of the photons in the bath influence on the rate of the irradiation and on the nanolaser quantum state.'' We can estimate the nonthermal reservoir correlation times. The correlation will be significant between the photons in the nonequilibrium reservoir that fly the distance of the order of the nanolaser size. So their correlation time: $t_{\rm cor} \sim l_{sp} /c\sim 10^{-15} s$. The smallest relaxation time of the nanolaser is of the order of $\tau _{a} \sim 10^{-14} s$. Because of the small number of the generated plasmons this time scale does not significantly change. So the relation $t_{\rm cor} \ll \tau _{a} $ remains true.

The system oscillates at optical frequencies that are much larger than the system-reservoir coupling constant as we discussed above (it should be noted that the rotating wave approximation is valid with accuracy $\sim \gamma ^{2} /\omega ^{2} $ where $\gamma $ is the interaction constant between system and reservoir and $\omega $ is the eigen frequency of the system, so called Bloch-Ziegert shift). Also we note that the nonequilibrium nature of the reservoir does not break this relation because of the small mean plasmon number. So the terms in the Hamiltonian that do not conserve energy may be neglected and so we again arrive to the rotating wave approximation. Another reason: the hottest photons in the nonequilibrium reservoir have the frequency of the order of the plasmon frequency and this is large frequency.
\bibliography{bibliography}
\end{document}